%% file: alkwai-arxiv19.tex
\documentclass[sigconf, nonacm=true]{acmart}

\usepackage{multirow}
\usepackage{algpseudocode}
\usepackage{algorithm}
\usepackage{graphicx}
\usepackage{subcaption}

\makeatletter
\algnewcommand{\LineComment}[1]{\Statex \hskip\ALG@thistlm \(\triangleright\) #1}
\makeatother

\begin{document}
\title{Making Recommendations from Web Archives for ``Lost'' Web Pages}

\author{Lulwah M. Alkwai, Michael L. Nelson, Michele C. Weigle}


\affiliation{%
  \institution{Department of Computer Science}
  \institution{Old Dominion University}
  \city{Norfolk}
  \state{Virginia 23529 USA}
}
\email{lalkw001@odu.edu, {mln, mweigle}@cs.odu.edu}


\begin{abstract}
When a user requests a web page from a web archive, the user will typically either get an HTTP 200
if the page is available,
or an HTTP 404
if the web page has not been archived.  This is because web archives are typically accessed by Uniform Resource Identifier (URI) lookup, and the response is binary: the archive either has the page or it does not, and
the user will not know of other archived web pages that exist and are potentially similar to the requested web page.
In this paper,
we propose augmenting these binary responses with a model for selecting and ranking recommended web pages in a Web archive.
This is to enhance both HTTP 404 responses and HTTP 200 responses by surfacing web pages in the archive that the user may not know existed.
First,
we check if the URI is already classified in DMOZ or Wikipedia.
If the requested URI is not found,
we use machine learning to classify the URI using DMOZ as our ontology and collect candidate URIs to recommended to the user.
The classification is in two parts,
a first-level classification and a deep classification.
Next,
we filter the candidates based on if they are present in the archive. 
Finally,
we rank candidates based on several features,
such as archival quality,
web page popularity,
temporal similarity,
and URI similarity.
We calculated the F\textsubscript{1} score for different methods of classifying the requested web page at the first level.
We found that using all-grams from the URI after removing numerals and the top-level domain (TLD) produced the best result with F\textsubscript{1}=0.59.
For the deep-level classification, we measured the accuracy at each classification level.
For second-level classification, the micro-average F\textsubscript{1}=0.30 and for third-level classification, F\textsubscript{1}=0.15.
We also found that 44.89\% of the correctly classified URIs contained at least one word that exists in a dictionary
and 50.07\% of the correctly classified URIs contained long strings in the domain.
In comparison with the URIs from our Wayback access logs, only 5.39\% of those URIs contained only words from a dictionary,
and 26.74\% contained at least one word from a dictionary.
These percentages are low and may affect the ability for the requested URI to be correctly classified.
\end{abstract}

\maketitle

\input{body}

\bibliographystyle{ACM-Reference-Format}
\bibliography{alkwai-arxiv19}

\end{document}

%% file: body.tex
\section{Introduction}
\begin{figure*}[t!]
\begin{subfigure}[c]{0.6\textwidth}
\frame{\includegraphics[width=\textwidth]{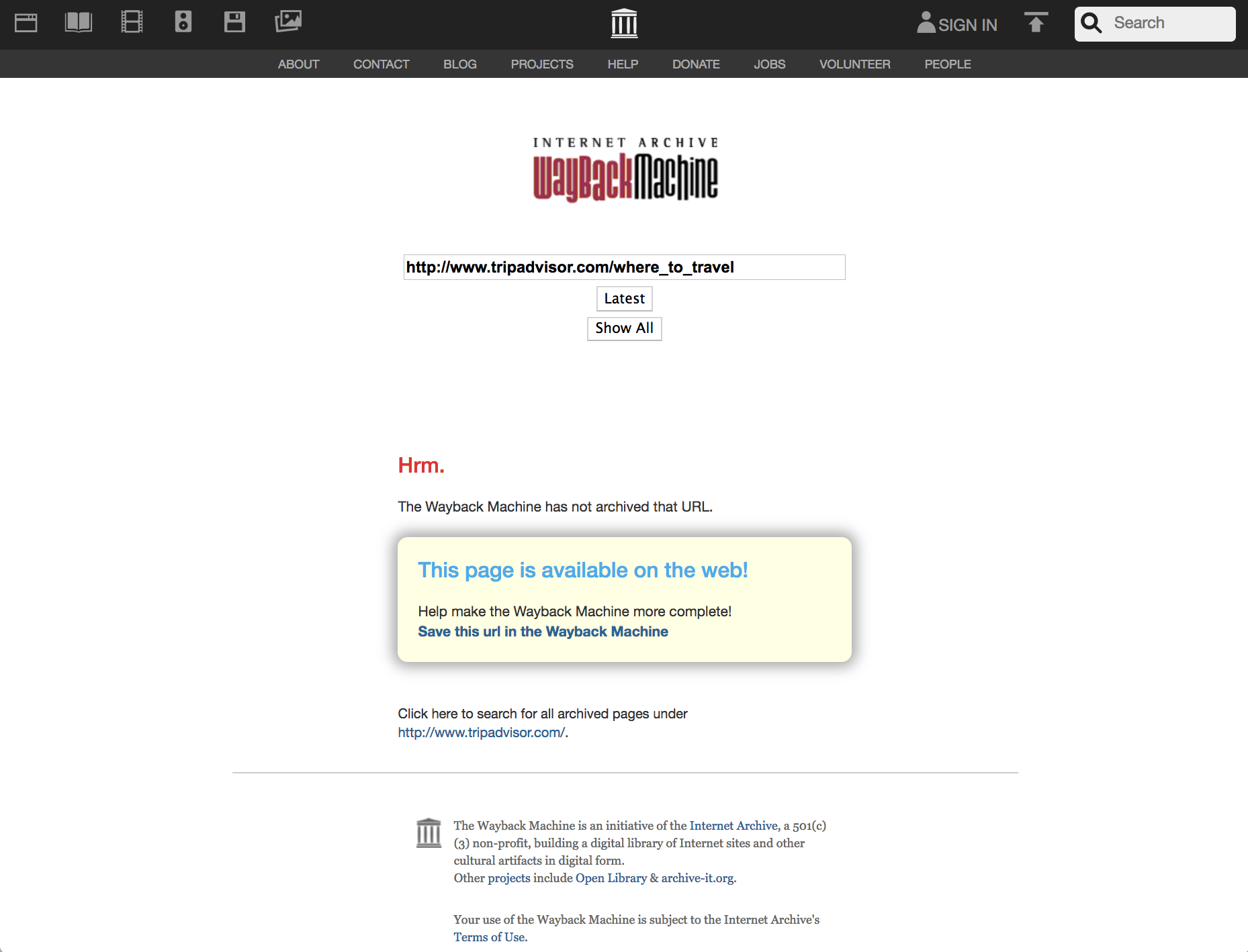}}
\caption{Response to the request \url{http://tripadvisor.com/where_to_travel} at the Internet Archive}\label{fig:wm}
\end{subfigure}
\hspace{2 cm}
\begin{subfigure}[c]{0.6\textwidth}
\frame{\includegraphics[width=\textwidth]{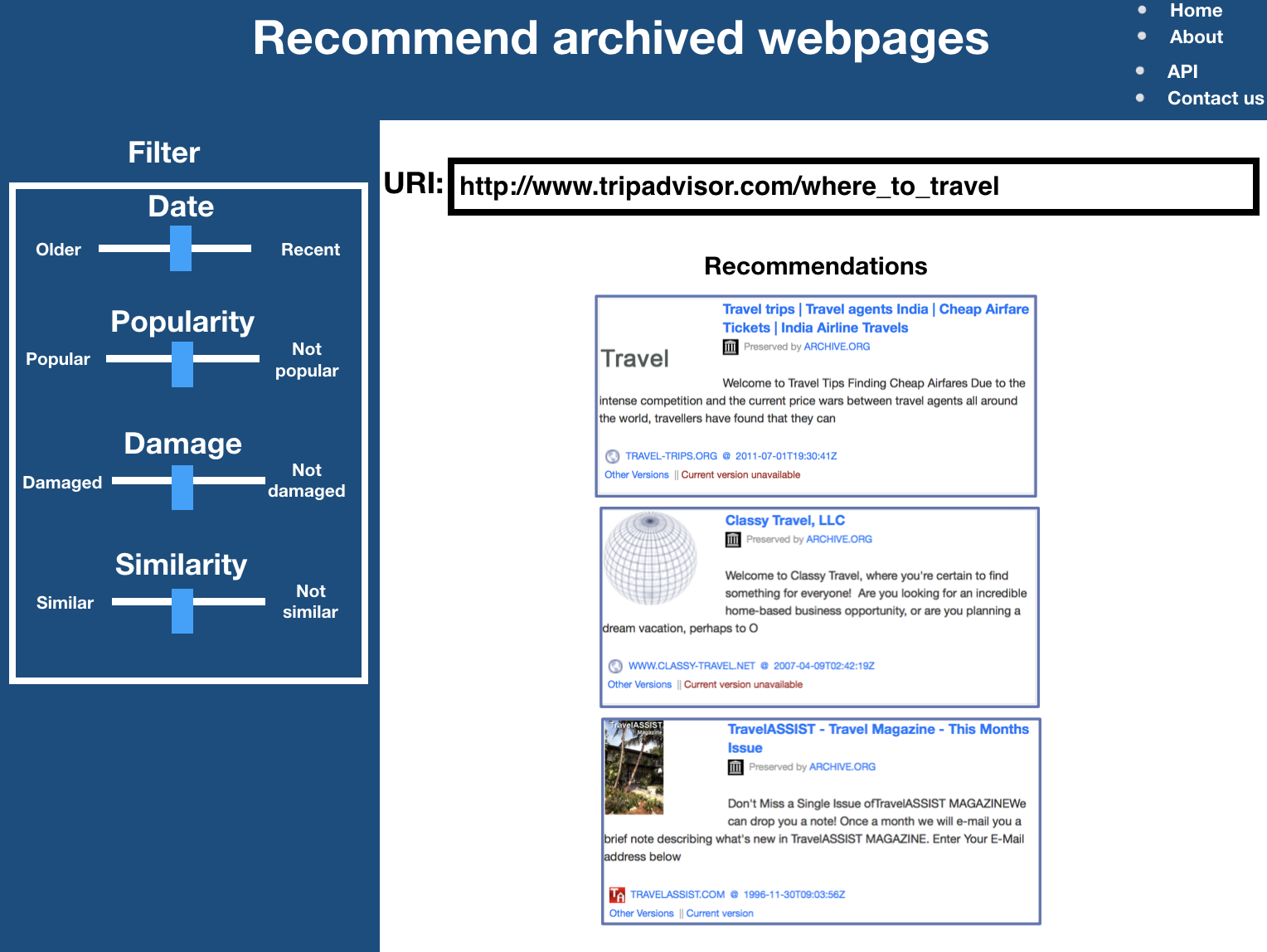}}
\caption{Proposed recommendations for the requested URI \url{http://tripadvisor.com/where_to_travel} displayed with  MementoEmbed \cite{Shawn_MementoEmbed} social cards}\label{fig:wm2}
\end{subfigure}
\caption{The actual response to the requested URI \url{http://tripadvisor.com/where_to_travel} (\ref{fig:wm}) and its proposed replacement (\ref{fig:wm2})}
\end{figure*}

Web archives are a window to view past versions of web pages.
The oldest and largest web archive,
the Internet Archive's Wayback Machine,
contains over 700 billion web objects \cite{Brewster_Kahle} .
But even with this massive collection,
sometimes a user requests a web page that the Wayback Machine does not have.
Currently,
in this case,
the user is presented with a message saying that the Wayback Machine does not have the page archived and a link to search for other archived pages in that same domain (Figure \ref{fig:wm}).
Our goal is to enhance the response from a web archive with recommendations of other archived web pages that may be relevant to the request.
For example,
Figure \ref{fig:wm2} shows a potential set of recommended archived web pages for the request in Figure \ref{fig:wm}.

One approach to finding related web pages is to examine the content of the requested web page and then select candidates with similar content.
However,
in this work,
we assume that the requested web page is neither available in web archives nor on the live web and thus is considered to be a ``lost'' web page.
This assumption reflects previous work showing that users often search web archives when they cannot find the desired web page on the live web \cite{alnoamany2014and} and that there are a significant number of web pages that are not archived \cite{ainsworth2011much, alkwai2017comparing}. 
Learning about a requested web page without examining the content of the page can be challenging due to little context and content available.
There are several advantages to using the Uniform Resource Identifier (URI) over using the content of the web page.
First,
in some cases the content of the URI is not available on the live Web or in the archive.
Second,
the URI may contain hints about the resource it identifies.
Third,
it is more efficient both in time and space to use the text of the URI only rather than to extract the content of the web page.
Fourth,
some web pages have little or no textual content,
such as images or videos,
so extracting the content will be not useful or even possible.
Fifth,
some web pages have privacy settings that do not permit them to be archived.


In this work we recommend similar URIs to a request by following five steps.
First,
we determine if the requested URI is one of the 4 million categorized URIs in DMOZ\footnote{The original DMOZ, \url{http://dmoz.org}, is out of service but we have archived versions locally.} or in Wikipedia via the Wikipedia API. 
If the URI is found, we collect candidates in the same category from DMOZ or Wikipedia and move to Step 4.
Second,
if the URI is not found we classify the requested URI based on a first-level of categorization.
Third,
we classify the requested URI to determine the deep categorization levels and collect candidates.
Fourth,
we filter candidates by removing candidates that are not archived.
Finally,
we filter and rank candidates based on several features,
such as archival quality,
web page popularity,
temporal similarity,
and URI similarity.
\section{Related Work}
There has been previous work on searching an archive without indexing it.
Kanhabua et al. \cite{kanhabua2016search} proposed a search system to support retrieval and analytics on the Internet Archive.
They used Bing to search the live web and then extracted the URLs from the results and used those as queries to the web archive.
They measured the coverage of the archived content retrieved by the current search engine and found that on page one of Bing results,
94\% are available in the Internet Archive.  Note that this technique will not find URLs that have been missing (HTTP status 404) long enough for Bing to have removed them from its index.  

Klein et al. \cite{klein2014moved} addressed a similar but slightly different problem by using web archives to recommend replacement pages on the live web.  They investigated four techniques for using the archived page to generate queries for live web search engines:
(1) lexical signatures,
(2) web page titles,
(3) tags,
and (4) link neighborhood lexical signatures.
Using these four methods helped to find a replacement for missing web pages.
Various datasets were used,
including DMOZ.
By comparing the different methods, they found that 70\% of the web pages were recovered using the title method.
The result increased to 77\% by combining the other three methods.
In their work,
the user will get a single alternative when a page is not found on the live Web.

Huurdeman et al. \cite{huurdeman2014finding, huurdeman2015lost} detailed their approach to recover pages in the unarchived Web based on the existence of links and anchors of crawled pages.
The data used was from the Dutch 2012 National Library of the Netherlands\footnote{\url{https://kb.nl/en}} (KB). 
Both external links (inter-server links),
which are links between different servers,
and site internal links (intra-server links),
which occur within a server,
were included in the dataset.
Their findings included that the archived pages show evidence of a large number of unarchived pages and web sites.
Finally,
they found that even with a few words to describe a missing web page,
they can be found within the first rank.

Classification is the process of comparing representations of documents with representations of labeled categories and computing similarity to find to which category the documents belong.
Baykan et al. \cite{baykan2009purely, baykan2011comprehensive} investigated using the URI to classify the web page and identify its topic.
They found that there is a relationship between classification and the length of the URI,
where the longer URI,
the better result.
They used different machine learning algorithms,
and the highest scores were achieved by the maximum entropy algorithm.
They trained the classifiers on the DMOZ dataset using all-grams method and tested the performance on Yahoo!,
Wikipedia,
Delicious,
and Google.
The classifier performed the best on the Google data,
with $F_1 = 0.87$.
We use Baykan et al.'s tokenization methods in Section \ref{sec:L1}.

Xue et al. \cite{xue2008deep} used text classification on a hierarchal structure.
They proposed a deep classification method,
where given a document,
the entire categories are divided into two kinds according to their similarity to the document,
related categories and unrelated categories.
They had two steps,
the search stage and the classification stage.
After the search stage ends a small subset of candidate categories in a hierarchy structure would be the result.
Then the output of the first stage would be the input of the second stage.
For the first search stage,
two strategies have been proposed,
document-based and category-based.
They either compared the requested document to each document in the dataset or compared it to all documents in a category.
Then term frequency (TF) and cosine similarity were used to find the top 10 documents.
For the second stage,
the resulting 10 category candidates are structured as a tree,
then the tree is pruned by removing the category if it has no candidate in it.
Three strategies are proposed to accomplish this step:
flat structure,
pruned top-down,
and ancestor-assistant.
They also used Na{\"i}ve Bayes as a classifier because of the large sample size and the speed desired.
They used 3-gram because of the close similarity between categories. 
As a dataset they used 1.3 million URIs from DMOZ and ignored the Regional and World categories.
For evaluation,
they used the Mi-F\textsubscript{1} score metric, which evaluates the performance for each level.
They found that the deep classification performs the highest of the three using the Mi-F\textsubscript{1} score,
where it resulted in a 77\% improvement over top-down based approach.
This work is the basis for the deep-level classification we perform (Section \ref{Sec:L2}).

Rajalakshmi et al. \cite{rajalakshmi2013web} proposed an approach where N-gram based features are extracted from URIs alone,
and the URI is classified using Support Vector Machines and Maximum Entropy Classifiers.
In this work,
they used the 3-gram features from the URI on two datasets:
2 million URIs from DMOZ and a WebKB dataset with 4K URIs.
Using this method on the WebKB dataset resulted in an increase of F\textsubscript{1} score by 20.5\%  compared to the related work \cite{kan2004web,kan2005fast,devi2007machine}.
Also,
using this method on DMOZ resulted in an increase of F\textsubscript{1} score by 4.7\% compared to the related work \cite{rajalakshmi2011naive,kan2005fast,baykan2009purely}.

One of the features we will use to rank the candidate URIs is the archival quality.
Archival quality refers to measuring memento damage by evaluating the impact of missing resources in a web page.
The missing resources could be text,
images,
video,
audio,
style sheet,
or any other type of resource on the web page.
Brunelle et al. \cite{brunelle2015not} proposed a damage rating algorithm to measure the relative value of embedded resources and evaluate archival success.
The algorithm is based on a URI's MIME type,
size,
and location of the embedded resources.
In the Internet Archive the average memento damage reduced from 0.16 in 1998 to 0.13 in 2013.

\section{Datasets}
In this work we use three datasets:
DMOZ,
Wikipedia,
and a set of requests to the Wayback Machine.
We use the DMOZ and Wikipedia datasets as ontologies to help classify the requested URI and generate candidate recommendations.
For evaluation, we use the Wayback Machine access logs as a sample of actual requests to a popular web archive.
We chose DMOZ because its web pages are likely to be found in the archive \cite{ainsworth2011much,alsum2014web}.
Wikipedia was chosen because new or recent web pages are found.
In this section we will describe each of the datasets.

\subsection{DMOZ}
\begin{figure*}
\frame{\includegraphics[width=5.2in]{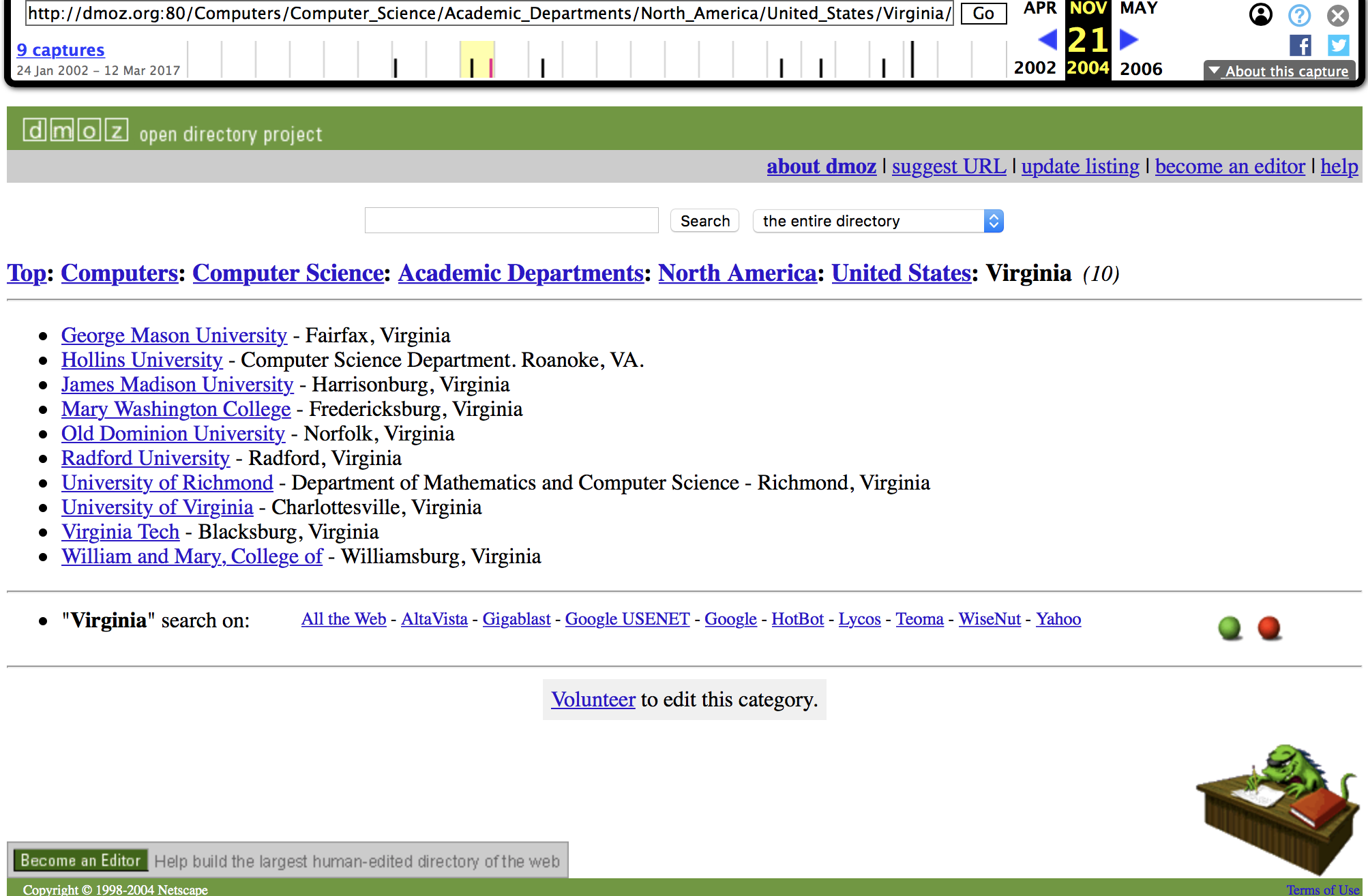}}
\caption{ODU main page found in DMOZ}\label{fig:dmoz}
\end{figure*}

DMOZ,
or the Open Directory Project (ODP),
was the largest human-edited directory of the Web.
DMOZ is considered a hierarchical classification in which each category may have sub-categories.
Each entry in the dataset contains the following fields:
category,
URI,
title,
and description.
For example an entry could be:
\url{Computers/Computer_Science/Academic_Departments/North_America/United_States/Virginia},
\url{http://cs.odu.edu/},
Old Dominion University,
and Norfolk Virginia,
as shown in Figure \ref{fig:dmoz}.

DMOZ was closed down on March 14, 2017.
We have archived 118 DMOZ files of the type RDF,
from 2001 to 2017.
Since we focus on English language web pages,
we first filtered out the World category.
Then,
we collect all entries that contain at least the URI and the category fields.
Next,
starting from the latest archived dataset,
we collected the entries that include a unique URI.
After that,
we converted all the URIs to Sort-friendly URI Reordering Transform (SURT)\footnote{\url{https://pypi.org/project/surt/}} format.
Table \ref{table_dmoz_cat} shows the number of collected entries and sub-categories for each category.
To be consistent with a similar work \cite{rajalakshmi2013web},
we filtered out the Regional, Netscape, Kids\_and\_Teens, and Adult categories.

\begin{table}
\caption{The number of entries for each category and the number of sub-categories in the DMOZ dataset}\label{table_dmoz_cat}
\begin{tabular}{|l|r|r|}
\hline
\multicolumn{1}{|c|}{\textbf{Category}}
&\multicolumn{1}{|c|}{\textbf{Num. URIs}}
&\multicolumn{1}{|c|}{\textbf{Num. sub-categories}}
\\ \hline
Regional
&2,348,257
&297,140
\\ \hline

Arts
&658,942
&57,959
\\ \hline

Society
&487,834
&36,259
\\ \hline

Business
&469,668
&22,465
 \\ \hline

News
&421,800
&2,581
\\ \hline

Computers
&297,789
&12,580
\\ \hline

Sports 
&278,706
&28,761
\\ \hline

Recreation
&261,005
&15,467
 \\ \hline

Shopping
&250,538
&7,393
\\ \hline

Science
&217,071
&17,212
\\ \hline

Adult
&197,141
&10,683
\\ \hline
 
Reference
&160,652
&13,077
\\ \hline
 
 Games  
&151,459
&20,233
\\ \hline

Health 
&149,648
&10,292
\\ \hline

Home
&81,059
&3,553
\\ \hline

Kids\_ands\_Teens
&63,333
&5,793
\\ \hline

Netscape
&27,223
&2,581
\\ \hline

Total
&6,522,125
&564,029
\\ \hline

\end{tabular}
\end{table}

Since we are going to gather recommendations from DMOZ,
we wanted to analyze the dataset.
We checked the top-level domains,
the depth of URIs,
if the URIs are on the live web,
and if URI patterns occur.

\paragraph{\textbf{Top-Level Domain}}~\\\label{tld_dmoz}
In this section we determine the diversity of the top-level domains (TLDs) in DMOZ.
Shown in Table \ref{table_dmoz_tld},
we found that 61.85\% of URIs are in the commercial top-level domain,
\url{.com},
followed by \url{.org},
\url{.net},
\url{.edu}.
Other top-level domains include \url{.ca},
\url{.it},
etc.
\begin{table}
\caption{Top-level domain analysis for DMOZ dataset}\label{table_dmoz_tld}
\begin{tabular}{|l|r|r|}
\hline
\multicolumn{1}{|c|}{\textbf{TLD}}
&\multicolumn{1}{|c|}{\textbf{Num. URIs}}
& \multicolumn{1}{c|}{\textbf{Percent}} \\ \hline
\textbf{com}
&4,034,276
&61.85\%
\\ \hline

\textbf{org}
&586,152
&8.99\%
 \\ \hline
 
\textbf{net}
&371,753
&5.70\%
 \\ \hline
 
\textbf{edu}
&224,539
&3.44\%
 \\ \hline
 
\textbf{gov}
&60,919
&0.93\%
 \\ \hline
 
\textbf{us}
&11,382
&0.17\%
 \\ \hline
 
\textbf{others}
&1,233,105
&18.90\%
\\ \hline

\textbf{Total}
&6,522,125
& 100\%
  \\ \hline
\end{tabular}
\end{table}

\paragraph{\textbf{Depth}}~\\
Here,
we want to know if the URIs we are recommending are only URIs of depth 0.
Note that depth 0 includes URIs ending with \url{/index.html} or \url{/home.html}.
The depth is measured after URI canonicalization\footnote{\url{https://pypi.org/project/surt/}}.
Shown in Table \ref{table_dmoz_depth} we found that 50.57\% of the URIs in DMOZ are depth 0 (i.e.,
top-level web pages).   
\begin{table}
\caption{Depth analysis for DMOZ dataset}\label{table_dmoz_depth}
\begin{tabular}{|c|r|r|}
\hline
\textbf{Depth}
& \multicolumn{1}{c|}{\textbf{Count}}
& \multicolumn{1}{c|}{\textbf{Percent}} \\ \hline

\textbf{0} 
&3,298,369
&50.57\%
\\ \hline

\textbf{1}
&1,134,874
&17.40\%
\\ \hline

\textbf{2}
&905,849
&13.89\%
 \\ \hline

\textbf{3+}
&1,183,033
&18.14\%
\\ \hline

\textbf{Total}
&6,522,125
& 100\%  \\ \hline

\end{tabular}
\end{table}

\paragraph{\textbf{Live Web}}~\\
As of November 2018,
we found that 86\% of the URIs in the DMOZ dataset are either live or redirect to live web pages.

\paragraph{\textbf{Patterns}}~\\
In this section we calculate the different URI patterns that occur in DMOZ.
Table \ref{DMOZ_patterns} shows the percentage of occurrence of the pattern in the hostname and the path.
We analyze the following patterns:

\begin{itemize}
\item \textbf{Long strings}.
Contains 10 or more contiguous letters.
We chose 10 because it is likely that at least two words are grouped together since the average English word is 5 letters long  \cite{pierce2012introduction,palmer1997trainable}.
Example: \url{http://radiotunis.com}.

\item \textbf{Long slugs}.
Contains 2 or more instances of 5 contiguous letters separated by non alphanumeric character.
Example: \url{http://www.arnosoftwaredev.blogspot.com/2005/01/sorting-algorithms-visualized.html}.

\item \textbf{Numbers}.
Example: \url{http://911.com}.

\item \textbf{Change in case}.
Example: \url{http://zeekoo.com/ZeeKooGids.php}.

\item \textbf{Query in the path}.
HTTP query string,
beginning with a ``?''.
Example:  \url{http://findagrave.com/cgi-bin/fg.cgi?page=gr&GRid=1795}.

\item \textbf{Port number in the hostname}.
Example: \url{http://www3.gencat.cat:81/justicia/justiterm/index.htm}.

\item \textbf{IP address in the hostname}.
Example: \url{http://63.135.118.69/}.

\item \textbf{Percent-encoding}.
Encoding to represent special characters in the URI.
Example: \url{http://tinet.cat/\%7ekosina}.

\item \textbf{Date string}.
Example: \url{http://elmundo-eldia.com/1999/08/29/opinion/1001023218.html}.

\end{itemize}

We found 42.65\% of the URIs contain long strings in the hostname and 20.01\% of the URIs contain numbers in the path.

\begin{table}
\caption{URI patterns present in DMOZ}
\label{DMOZ_patterns}
\begin{tabular}{|l|r|r|}
\hline
\multicolumn{1}{|c|}{\textbf{Pattern}}

&\multicolumn{1}{|c|}{\textbf{\% in hostname}}
&\multicolumn{1}{|c|}{\textbf{\% in path}}
\\ \hline

\textbf{Long strings}
&42.65\%
& 13.21\%\\ \hline

\textbf{Long slugs}
&10.85\%
&7.82\%\\ \hline

\textbf{Numbers}
&4.37\%
&20.01\% \\ \hline

\textbf{Change in case}
&0.36\% 
&8.18\%\\ \hline

\textbf{Query}
&-
&4.72\%\\ \hline

\textbf{Port number}
&0.11\%
&-\\ \hline

\textbf{IP address}
&0.07\% 
&-\\ \hline

\textbf{Percent-encoding}
&0\% 
&0.50\%\\ \hline

\textbf{Date}
&0\%
&0.43\% \\ \hline

\end{tabular}
\end{table}

\subsection{Wikipedia} 
\begin{figure*}
\includegraphics[width=6in]{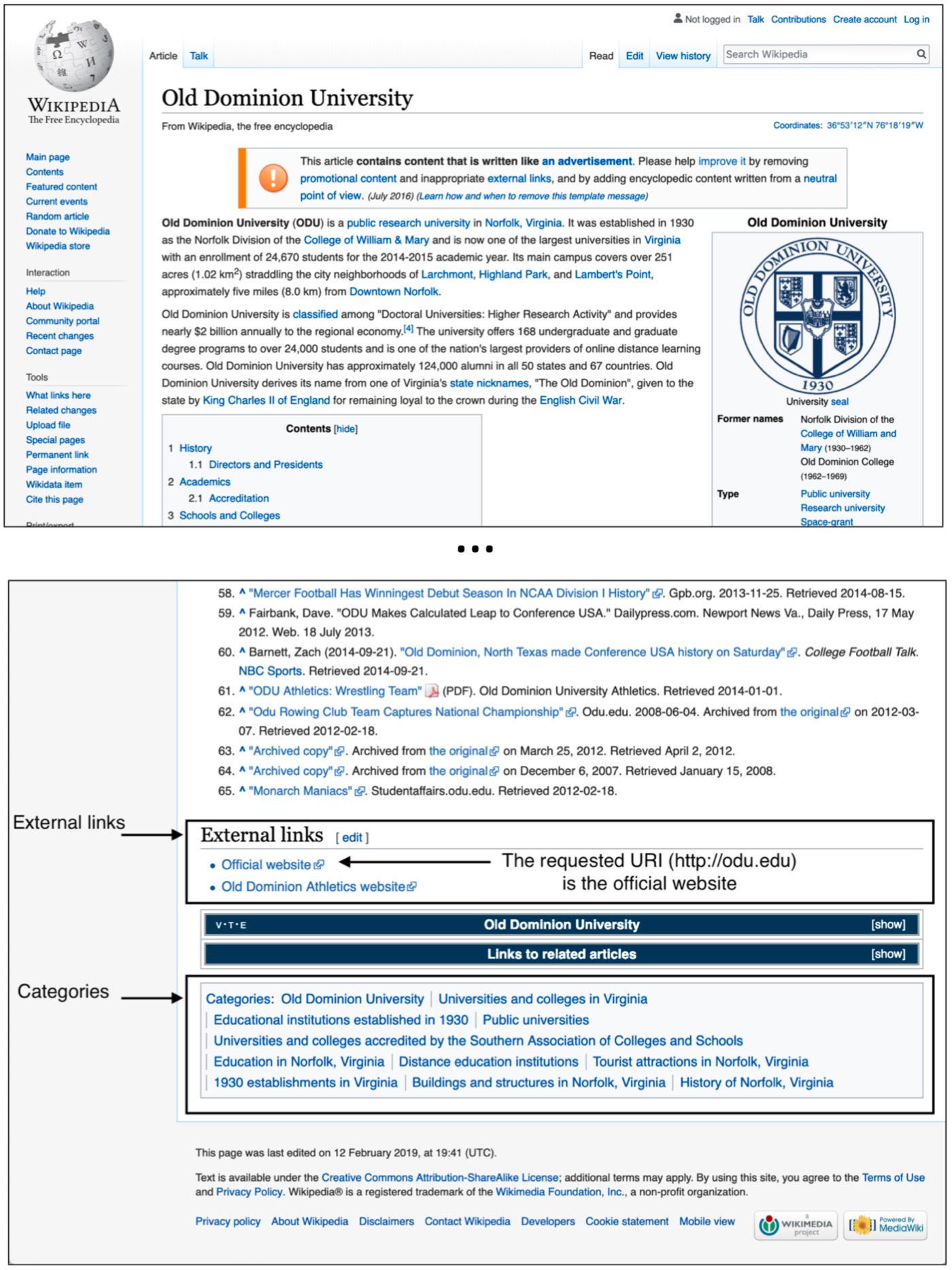}
\caption{Searching for the request \url{http://odu.edu} in Wikipedia resulted in finding the Wikipedia web page \url{https://en.wikipedia.org/wiki/Old_Dominion_University} that contains the requested URI as the official website in the external link section. We use other web pages in the same categories (at the end of the page) as candidate web pages.}\label{fig:wiki_odu}
\end{figure*}

Wikipedia is a web-based encyclopedia,
launched in 2001 \cite{wiki_history} and available in 304 languages \cite{wiki_lang}.
It contains articles that are categorized and most also contain a list of external links.
For instance,
the article shown in Figure \ref{fig:wiki_odu} is categorized as
\textit{Old Dominion University, Universities and colleges in Virginia, Educational institutions established in 1930}, etc.
and contains two external links at the end of the article.
If the entity described in the article has an official website, then
it will be linked as the ``Official website'' in the list of external links.
We use Python Wikipedia packages \cite{jonathan-goldsmith,martin-majlis} to extract the information needed.

\subsection{Wayback Machine}\label{sectionWM}
The Wayback Machine server access logs contain real requests to the Internet Archive's Wayback Machine \cite{tofel2007wayback}.
The requests are from 295 noncontiguous days between 2011-01-01 to 2012-03-02.
A sample of this dataset was used for evaluation.
This dataset has been used in other work \cite{alnoamany2013access, alnoamany2016using}.

Each request (line) contains the following information:
Client IP,
Access Time,
HTTP Request Method,
URI,
Protocol,
HTTP Status Code,
Bytes Sent,
Referring URI,
User-Agent.

In our work,
we will use a sample from the requests made on Feb 8, 2012,
similar to data selected in AlNoamany et al. \cite{alnoamany2013access}.
There were 49,026,577 requests on that day.
Before collecting a sample to use,
we performed several filtering steps.
First,
we filtered out any requests that did not result in an HTTP 200 status code.
We also filtered out any requests with an invalid URI format or extension,
non-HTML URIs,
an IP address as the domain,
or a ccTLD from a non-English speaking country.
In addition, we filtered out requests that resulted in HTML with a non-English HTML language code.
This filtering left 732,130 unique URIs.

\section{Algorithm}
Our recommendation algorithm,
shown in Algorithm \ref{recommending_algo},
is composed of four main steps,
each of which will be described in more detail in the following subsections.
As per the current method of searching a web archive,
the user provides a requested URI and optionally a desired datetime.

Our goal is to provide recommendations for other archived web pages based on the requested URI,
which we assume is ``lost'',
neither available on the live web nor archived.
The first step is to obtain a first-level classification of the URI,
using DMOZ or Wikipedia.
This would result in a high-level category for the URI,
such as ``Computers'',
``Business'',
etc.
similar to those in Table \ref{table_dmoz_cat}.
We then use machine learning techniques to obtain a deeper categorization,
such as ``\url{Computers/Computer_Science/Academic_Departments/
North_America_United_States/Virginia}''.
Once this categorization is obtained,
we can collect candidates from other URIs in the same category in DMOZ or Wikipedia.
Then we filter out any candidates that are not archived and finally rank and recommend candidates based on several features,
such as archival quality,
web page popularity,
temporal similarity,
and URI similarity.
\begin{algorithm}
  \caption{Algorithm for recommending archived web pages using only the URI}\label{recommending_algo}
  \begin{algorithmic}[0]

  \LineComment {Step 1: Classify the URI ($level one$)}  
  \Function {Classify\_URI\_level\_one} {$requested\_URI$}
  	    		\State Tokenize (requested\_URI)
    			\State ML (requested\_URI)  
  \EndFunction
  
   \State
  \LineComment {Step 2: Deep classify the URI ($deep-levels$)}
  \Function {Classify\_URI\_deep\_levels} {$requested\_URI$}
  	  		\State Index\_dataset\_by\_category ()
			\State  Cosine\_similarity ($requested\_URI$)
	  		\State  Get\_top\_N\_candidates ($Candidates$)
	  		\State  Create\_and\_prune tree ($Candidates$)
	  		\State  ML ($Candidates$)
  \EndFunction
  
   \State
  \LineComment {Step 3: Filter candidates}
   \Function {Archived}{$Candidates$}
  		 \For{$Candidates$} \If{Candidate is archived}\State Archived\_Candidates=Candidate\EndIf
		\EndFor
  \EndFunction
  
   \State
  \LineComment {Step 4: Score and rank candidates}
   \Function {Rank}{$Archived\_Candidates$}
     \State Score ($Archived\_Candidates$)
     \State Get\_top\_N\_candidates ($Archived\_Candidates$)
   \EndFunction
    
     \State
  \LineComment {Main Function}
  \Function {Recommending\_Archived\_Web\_Pages}{$requested\_URI$}
   \If{requested\_URI not in a\_classified\_ontology}
			\State  \Call{Classify\_URI\_level\_one}{$requested\_URI$}\Comment{Step 1}
    			\State  \Call{Classify\_URI\_deep\_levels}{$requested\_URI$}\Comment{Step 2}
    \EndIf
    \State \Call{Collect\_All\_Candidates}{$requested\_URI$}
    \State  \Call{Archived}{$Candidates$}\Comment{Step 3}
    \State \Call{Rank}{$Archived\_Candidates$} \Comment{Step 4}
    \EndFunction
  \end{algorithmic}
\end{algorithm}
\subsection{Check Ontologies}
The first step is to determine if the requested URI is already present and categorized in 
DMOZ or Wikipedia.
Using DMOZ is straightforward;
we check if the URI exists in DMOZ or not.
However,
in Wikipedia we check if the requested URI is the official web site (by searching for the keyword ``official website'') and is categorized.
For example,
if the requested URI was 
\url{http://odu.edu},
we use the URI to find a related Wikipedia web page.
In this example we find that the Wikipedia web page 
\url{https://en.wikipedia.org/wiki/Old_Dominion_University} mentions \url{http://odu.edu} as the official website.
Then we collect the categories that this web page belongs to, such as 
\textit{Old Dominion University, Universities and colleges in Virginia, Educational institutions established in 1930}, etc.
Then we collect as candidates all of the official web pages that these categories contain.

To test how often this option might be available,
we used the Wayback Machine access logs (Section \ref{sectionWM}).
From the filtered set,
we found 13.17\% URIs in DMOZ or Wikipedia.  

\subsection{Step 1: First-Level Classification}\label{sec:L1}
For a request that did not appear in an ontology,
we will classify it using only the tokens from the URI.
We test three different methods of tokenization.
First,
we use URI tokens that are split by non-alphanumeric characters.
Second,
we use all-grams from the tokens.
Third,
we use all-grams from the URI.

\subsubsection{\textbf{Tokenize the URI}}~\\
To classify the URI,
we need to extract meaningful keywords,
or tokens,
from the URI.
We adopt the three methods proposed by Baykan et al. \cite{baykan2011comprehensive}.

\begin{itemize}
\item \textbf{Tokens}
The URI is split into potentially meaningful tokens.
The URI is converted to lower-case and then split into tokens using any non-alphabetic character as a delimiter.
Finally,
the ``http'' (or ``https'') token is removed,
along with any resulting token of length 2 or less.

\item \textbf{All-grams from tokens}
The URI tokens are converted to all-grams.
We perform the tokenization as above and then generate all-grams on the tokens by combining 4-,
5-,
6-,
7-,
and 8-grams of the combined tokens.

\item \textbf{All-grams from the URI}
The URI is converted to all-grams without tokenizing first.
Any punctuation and numbers are removed from the URI,
along with ``http'' (or ``https'').
Then the result is converted to lowercase.
Finally,
the all-grams are generated by combining the 4-,
5-,
6-,
7-,
and 8-grams of the remaining URI characters.
\end{itemize}

An example of the different tokenization methods is shown in Table \ref{table_tokenize_ex}.
Using these methods we also examine removing the TLDs from the URIs,
removing numbers,
and removing stop words (Section \ref{ClassifyrequestusingMA}).

\begin{table}
\caption{Tokenizing the URI \url{https://odu.edu/compsci} using different methods \cite{baykan2011comprehensive}}\label{table_tokenize_ex}

\begin{tabular}{|l|l|}
\hline
\multicolumn{1}{|c|}{\textbf{Method}} 
& \multicolumn{1}{c|}{\textbf{Result}}                                                                                                                                                                                                                                                                                                                                                \\ \hline

Tokens
& odu, edu, compsci\\ \hline

All-grams from tokens
& \begin{tabular}[c]{@{}l@{}}odu, edu, comp, omps,\\ mpsc, psci, comps, ompsc,\\ mpsci, compsc, ompsci, compsci\end{tabular}\\ \hline

\begin{tabular}[c]{@{}l@{}}All-grams from URI\\ (\url{http://odu.edu/compsci})\end{tabular}
& \begin{tabular}[c]{@{}l@{}}odue, dued, uedu, educ,duco,\\ ucom, comp, omps, mpsc, psci,\\ odued, duedu, ueduc, educo,\\ ducom, ucomp, comps, ompsc,\\ mpsci, oduedu, dueduc, ueduco,\\ educom, ducomp, ucomps, compsc,\\ ompsci, odueduc, dueduco,\\ ueducom, educomp, ducomps,\\ ucompsc, compsci, odueduco,\\ dueducom, ueducomp, educomps,\\ ducompsc, ucompsci\end{tabular} \\ \hline

\end{tabular}
\end{table}

To determine the best tokenization method, as a baseline
we tested the classification of tokens on the DMOZ dataset,
using machine learning.
We took the DMOZ dataset and created a 10-fold cross-validation set,
using 90\% for training and 10\% for testing.
We employed a Na{\"i}ve Bayes classifier to take tokens and return the top-level category.
Na{\"i}ve Bayes was selected because of its simplicity that assumes independence between the features.
In the testing dataset we filtered out URIs that contain tokens not seen in the training set,
as was also done in related work \cite{baykan2011comprehensive}.

We measured the F\textsubscript{1} score to evaluate the different tokenization methods.
Table \ref{table_dmoz_classifiying1} shows the result of our evaluation.
In addition to the base tokenization methods described above,
we also tested the following alternatives for each method:
\begin{itemize}
\item remove TLD before tokenization
\item remove TLD and numbers before tokenization
\item remove TLD, numbers, and stop words before tokenization
\end{itemize}
The stop words were based on a set of stop words in the Natural Language Toolkit (NLTK)\footnote{\url{https://nltk.org/}}.
We found that using the all-grams from the URI after removing the TLD and numbers had the highest F\textsubscript{1} score,
which was comparable to results obtained in related work \cite{rajalakshmi2013web}.
We use this method of tokenization going forward.

\begin{table}
\small{
\caption{Classifying at the first-level, comparing F\textsubscript{1} score, Micro average, and Macro average for DMOZ dataset using different methods}\label{table_dmoz_classifiying1}
\begin{tabular}{|c|c|r|r|r|}
\hline
\multicolumn{2}{|c|}{\textbf{Method}}
& \multicolumn{1}{c|}{\textbf{\begin{tabular}[c]{@{}c@{}}F\textsubscript{1} score\end{tabular}}} 
& \multicolumn{1}{c|}{\textbf{\begin{tabular}[c]{@{}c@{}}Micro\\ average\end{tabular}}}
& \multicolumn{1}{c|}{\textbf{\begin{tabular}[c]{@{}c@{}}Macro\\ average\end{tabular}}} \\ \hline

\multirow{4}{*}{\textbf{Tokens}} 
& \textbf{All URI tokens}
&0.39
&0.45
&0.31
\\ \cline{2-5} 

& \textbf{\begin{tabular}[c]{@{}c@{}}URI tokens,\\ without TLD\end{tabular}}
&0.35
&0.40
&0.28
\\ \cline{2-5} 

& \textbf{\begin{tabular}[c]{@{}c@{}}URI tokens,\\ without TLD\\ and numbers\end{tabular}}
&0.40
&0.45
&0.32
\\ \cline{2-5} 

& \textbf{\begin{tabular}[c]{@{}c@{}}URI tokens,\\ without TLD\\ and stop words\end{tabular}}
&0.39
&0.43
&0.30
\\ \hline
\multirow{4}{*}{\textbf{\begin{tabular}[c]{@{}c@{}c@{}}All-gram\\ from\\ tokens\end{tabular}}}
& \textbf{All URI tokens}
&0.51
&0.53
&0.45
\\ \cline{2-5} 

& \textbf{\begin{tabular}[c]{@{}c@{}}URI tokens,\\ without TLD\end{tabular}}
&0.51
&0.53
& 0.46
\\ \cline{2-5} 

& \textbf{\begin{tabular}[c]{@{}c@{}}URI tokens,\\ without TLD\\ and numbers\end{tabular}}
&0.51
&0.52
&0.47
\\ \cline{2-5} 

& \textbf{\begin{tabular}[c]{@{}c@{}}URI tokens,\\ without TLD\\ and stop words\end{tabular}}
&0.50
&0.52
& 0.46
\\ \hline

\multirow{4}{*}{\textbf{\begin{tabular}[c]{@{}c@{}c@{}}All-grams\\ from\\ URI\end{tabular}}}
& \textbf{All URI tokens}
&0.56
&0.55
&0.48
\\ \cline{2-5} 

& \textbf{\begin{tabular}[c]{@{}c@{}}URI tokens,\\ without TLD\end{tabular}}
&0.55
&0.59
&0.46
\\ \cline{2-5} 

& \textbf{\begin{tabular}[c]{@{}c@{}}URI tokens,\\ without  TLD\\ and numbers\end{tabular}}
&\textbf{0.59}
&\textbf{0.62}
&\textbf{0.61}
\\ \cline{2-5} 

& \textbf{\begin{tabular}[c]{@{}c@{}}URI tokens,\\ without TLD\\ and stop words\end{tabular}}
&0.55
&0.60
&0.47
\\ \hline
\end{tabular}
}
\end{table}

\subsubsection{\textbf{Classify the URI using Machine Learning}}\label{ClassifyrequestusingMA}~\\
Now that we have determined the best tokenization method,
we will apply this for future requests.
We trained the Na{\"i}ve Bayes classifier on the entire DMOZ dataset and this will be used for classification as the baseline at the first-level.
We take the requested URI,
remove the TLD and numbers,
and then perform the all-gram from URI tokenizations described in the previous section.
These resulting all-grams are used in the the classifier to produce a first-level classification.

\subsection{Step 2: Deep-Level Classification}\label{Sec:L2}
In this step we want to classify the requested URI \url{http://cs.odu.edu/compsci} to a hierarchal deep classification such as \url{Computers/Computer_Science/Academic_Departments/
North_America_United_States/Virginia}.
Known methods to determine hierarchal deep classification are the big-bang approach and the top-down approach \cite{sun2001hierarchical}.
Neither method is ideal with a large number of hierarchies and may result in error propagation.
For this reason we adopt the method by Xue et al. \cite{xue2008deep}, but as opposed to this work, we are limited to the URI only and do not have the documents or any supporting details.

\begin{enumerate}

\item\textbf{Index dataset}.
In preparation to compute similarity between the requested URI and the category entries,
we index DMOZ by category,
creating a list of all URIs in each of the DMOZ deep-level categories.

\item\textbf{Cosine similarity}.
We compute the cosine similarity between the tokenized requested URI and the tokenized URIs and their titles and description,
in each category.
In this step each category of the index will get a similarity score to the requested URI,
which is the average similarity to all entries in that category.

\item\textbf{Collect N candidates}.
Next we select the top 10 candidate categories with the highest similarity score,
similar to related work \cite{xue2008deep}.

\item\textbf{Prune tree}.
Each candidate category could be a leaf node or an internal node.
We create a hierarchical tree and then prune it to get the final list of candidates that we can use machine learning to classify.
First,
we create a tree from the candidates by starting from the first node and then going down until all 10 candidates are presented,
as shown in Figure \ref{tree}.
Next,
in order to enhance the classification,
the tree is pruned based on the ancestor assistance strategy.
The ancestor assistance strategy includes the ancestors of a node if there are no common ancestors with another candidate,
as shown in Figure \ref{prune}.
\begin{figure}
\begin{subfigure}[b]{0.4\textwidth}
\includegraphics[ width=3.2in]{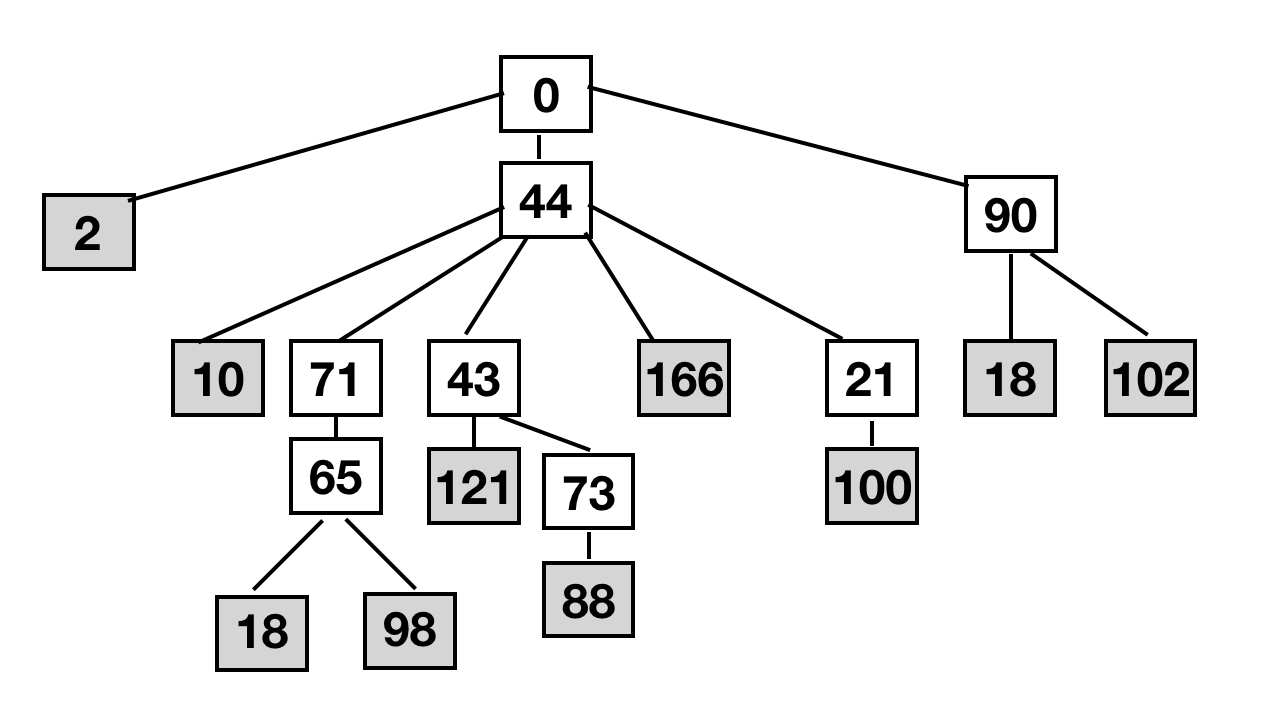}
\caption{Create hierarchical tree from the 10 candidate categories (the candidate categories are highlighted). The numbers represent the category ID}
\label{tree}
\end{subfigure}

\begin{subfigure}[b]{0.4\textwidth}
\includegraphics[width=3.2in]{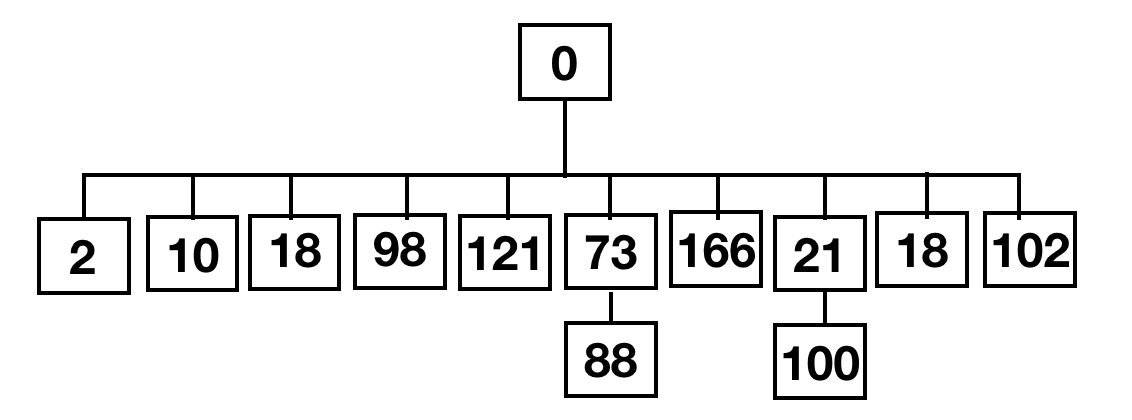}
\caption{Pruned tree using ancestor assistance strategy. The parents of nodes 88 and 100 are included because they have no shared ancestor with other candidates}
\label{prune}
\end{subfigure}
\caption{The process of pruning a hierarchical tree using ancestor assistance strategy \cite{xue2008deep}}\label{fig:prune_t}
\end{figure}

\item\textbf{Classify}.
To choose a single classification from the pruned tree we classify the requested URI based on two methods,
using 3-gram tokens and all-grams.
The 3-gram method had the best result when comparing documents \cite{xue2008deep},
however in our work we compare URI tokens,
so we expect the  all-gram method to perform better.
\end{enumerate}

\subsection{Steps 3, 4: Filter, Rank and Recommend}
Step 3 in our algorithm is to ensure that all recommendations come from a web archive.
We take the candidates from Step 2 and remove any that are not archived.
We use MemGator \cite{alam2016memgator} to determine this.
In Step 4,
we rank and recommend the remaining candidates based on temporal similarity ($t$),
web page popularity ($p$),
URI similarity ($s$),
and archival quality ($q$).
Our final list of recommended web pages will be ranked based on Equation \ref{eq1},
where \textit{w\textsubscript{t}+w\textsubscript{p}+w\textsubscript{s}+w\textsubscript{q}}=1.0 and specify the weights given to each of the features.
\begin{equation}\label{eq1}
score= w_tt + w_pp + w_ss + w_qq
\end{equation}

\subsubsection{\textbf{Temporal similarity}}\hfill\\
Temporal similarity refers to how close the available candidate web page's Memento-Datetime \cite{memento:rfc} is to the requested URI.
This is shown in Equation \ref{eq_t},
where $r_d$ is the request datetime,
$c_d$ is the candidate datetime,
$u_d$ is the current datetime,
and $e_d$ is the earliest datetime.
The earliest datetime is considered 1996,
because it was when archiving the Web started\footnote{\url{https://archive.org/about/}}.
\begin{equation}\label{eq_t}
t = \frac{|r_d- c_d|}{u_d - e_d}
\end{equation}

\subsubsection{\textbf{Web page popularity}}\hfill\\
We use how often the web page has been archived and the domain's popularity as determined by Alexa\footnote{\url{https://alexa.com}} as an approximation for the web page's popularity.
Our popularity measure $p$ is given in Equation \ref{eq_p},
where $a$ is the Alexa Global Ranking of the requested domain,
$x$ is the lowest ranked domain in Alexa,
$n$ is the number of times the URI has been archived,
and $m$ is the number of times Alexa's top-ranked web site has been archived.
\begin{equation}\label{eq_p}
p = \frac{(|\frac{log a}{log x}-1|+\frac{\log{n}}{\log{m}})}{2}
\end{equation}
We set $x$ to 30,000,000 as it is the current lowest ranking in Alexa,
and we set $m$ to 538,300,
the number of times that \url{http://google.com},
the top-ranked Alexa web page,
has been archived.

\subsubsection{\textbf{URI similarity}}\hfill\\
We measure the similarity of requested URI tokens and candidate URI tokens using Jaccard similarity coefficient (Equation \ref{eq_s}).
\begin{equation}\label{eq_s}
s = \frac{|A \cap B|}{|A|+|B|-|A \cap B|}
\end{equation}

\subsubsection{\textbf{Archival quality}}\hfill\\
Archival quality refers to how well the page is archived.
We use Memento-Damage \cite{Erika_Siregar} to calculate the impact of missing resources in the web page.
We calculate archival quality in Equation \ref{eq_q},
where $d$ is the damage score calculated from Memento-Damage.
\begin{equation}\label{eq_q}
q= |d-1|
\end{equation}

\section{Example}
Here we present an example of a request and the resulting recommendations. We request
\url{http://odu.edu/compsci} with the date of March 1, 2014.
This URI is not classified in DMOZ or in Wikipedia,
so we use machine learning and classify it to \url{Computers/Computer_Science/Academic_Departments/North_America/United_States/Virginia}.
Then we collect all the candidates from DMOZ:
\begin{itemize}
\item \url{http://cs.gmu.edu}
\item \url{http://cs.odu.edu}
\item \url{http://cs.virginia.edu}
\item \url{http://cs.vt.edu}
\item \url{http://wm.edu/as/computerscience/?svr=web}
\item \url{http://radford.edu/content/csat/home/itec.html}
\item \url{http://cs.jmu.edu}
\item \url{https://php.radford.edu/~itec}
\item \url{http://mathcs.richmond.edu}
\item \url{http://hollins.edu/academics/computersci}
\end{itemize}
Using equal weights ($w_t = w_p = w_q$) for our ranking equation,
the top three ranked candidates are:

\begin{enumerate}
\item \url{https://web.archive.org/web/20140226090846/http://cs.odu.edu:80/},
score= 0.87
\item \url{https://web.archive.org/web/20140208043915/http://cs.virginia.edu/},
score= 0.75
\item \url{https://web.archive.org/web/20140223213510/http://cs.jmu.edu/},
score= 0.73
\end{enumerate}

\section{Evaluation and Results}
\begin{figure}
\includegraphics[width=3.2in]{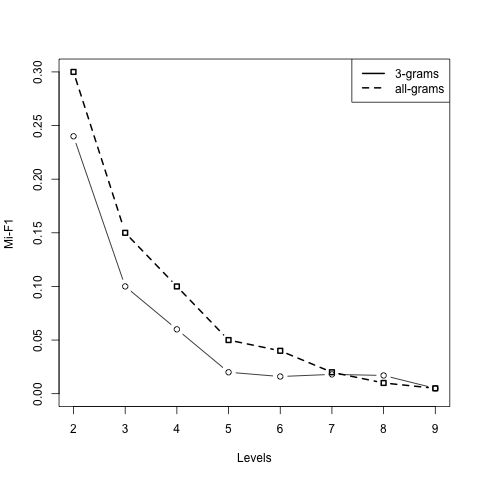}
\caption{Performance on classifying to different levels using 3-gram and all-gram}
\label{chart_deep_3}
\end{figure}

First,
we evaluate how well our deep classification method works (Step 3).
To test this step we use 10\% of the DMOZ dataset for testing and the rest for training.
We assume that level one categorization is already predicted in Step 1.
We evaluate the performance by determining if we classified each level correctly.
For example,
if a URI is actually in the category \url{c1/c2/c3},
then for level two evaluation,
we check if we predicted \url{c1/c2}.
For each level we calculate the Micro-average F\textsubscript{1} (Mi-F\textsubscript{1}) score.
In Figure \ref{chart_deep_3},
we show the Mi-F\textsubscript{1} score of each level using 3-gram cosine similarity.
The highest level in our results was 0.2 compared to 0.8 in the related work \cite{xue2008deep},
but that is due to using only the requested URI as the testing data and the URI and title and category as training, as opposed to using the text of the full document as in \cite{xue2008deep}.  
This shows that using only the tokens from the URI is not enough for deep classification.
Because of limited information,
we also show the result of testing the same method using all-gram cosine similarity.
We found that the results are better,
however it is still considered low compared to related work.

Some features could affect the URI classification.
We investigated the relationship between the depth of the URI and classification.
Table \ref{table_depth_class} shows the URI depth and the percentage of the correctly classified URIs.
We only considered URIs to be correctly classified if they were correct to the deepest level.
We found that 63.45\% of the correctly classified URIs are of depth 0.

\begin{table}
\caption{URI depth and percentage of correctly classified URIs}\label{table_depth_class}
\begin{tabular}{|c|r|}
\hline
\textbf{Depth} &\textbf{Percent} \\ \hline
0     & 63.45\%      \\ \hline
1     & 16.96\%      \\ \hline
2     &  13.48\%     \\ \hline
3     &  3.77\%     \\ \hline
4     &  1.47\%     \\ \hline
5+     &  0.86\%     \\ \hline
\end{tabular}
\end{table}

Next,
we check if the words in the URIs are in a dictionary (after removing the TLD).
We use the enchant English dictionary\footnote{\url{https://pypi.org/project/pyenchant/}} and wordninja\footnote{\url{https://pypi.org/project/wordninja/}} to split compound words.
For example,
the URI \url{http://mickeymantlebaseballcards.net} is split to mickey,
mantle,
baseball,
and cards.
We found that 36.92\% of the correctly classified URIs contain only words from a dictionary,
and 44.89\% of the correctly classified URIs contain at least one word from a dictionary.

An ideal structure of the URI is that it contains long strings that will have more semantics.
We are trying to identify a ``slug'',
which is the part of a URI that contains keywords or the web page title.
An example of a slug is the path in \url{https://cnn.com/2017/07/31/health/climate-change-two-degrees-studies/index.html}.
The slug in the URI is readable,
and we can identify what the web page is about.
We evaluate the existence of long strings in the correctly classified URIs.
We assume that the average length of an English word is 5 \cite{pierce2012introduction,palmer1997trainable} and anything greater is considered a long string.

Overall,
we found that 41.58\% of the sampled URIs contain long strings in the domain,
for example,
\url{http://timesonline.co.uk/tol/sport/cricket/}.
Also,
we found that 89.47\% of the sampled URIs contain long strings in the path,
for example,
\url{http://medlineplus.nlm.nih.gov/medlineplus/parkinsonsdisease.html}.
When analyzing the correctly classified URIs,
we found that 50.07\% of the correctly classified URIs contain long strings in the domain.
Also,
we found that 13.45\% of the correctly classified URIs contain long strings in the path.
Words can be separated by delimiters in the domain or path.
We found that 9.91\%  of the correctly classified URIs contain words separated by delimiters in the domain,
for example,
\url{http://vintage-poster-art.com/}.
We also found that 6.97\%   of the correctly classified URIs contain separated words by delimiters in the path,
for example,
\url{http://seaworldparks.com/en/buschgardens-williamsburg/}.

In addition,
we wanted to investigate the effect of the category on correct classification.
As shown in Table \ref{table_cat_per},
we found that 15.32\% of the correctly classified URIs were in the ``Society'' first-level category.
We also found that none of the correctly classified URIs were in ``News''.
We found that in the ``News'' category in DMOZ,
there is a level two subcategory ``Online\_Archive'' that contains 95\% of the ``News'' URIs and repeats several subcategories inside ``News''.
This caused errors in our classification.

\begin{table}
\caption{Percentage from the correctly classified URIs for each category}\label{table_cat_per}
\begin{tabular}{|l|r|r|}
\hline
\multicolumn{1}{|c|}{\textbf{Category}} & \multicolumn{1}{c|}{\textbf{count}} & \multicolumn{1}{c|}{\textbf{Percent}} \\ \hline
\textbf{Society}                        & 459                                 & 15.32\%                               \\ \hline
\textbf{Arts}                           & 401                                 & 13.38\%                               \\ \hline
\textbf{Shopping}                       & 355                                 & 11.85\%                               \\ \hline
\textbf{Recreation}                     & 331                                 & 11.05\%                               \\ \hline
\textbf{Sports}                         & 291                                 & 9.71\%                               \\ \hline
\textbf{Home}                           & 288                                 & 9.61\%                                \\ \hline
\textbf{Reference}                      & 238                                 & 7.94\%                                \\ \hline
\textbf{Computers}                      & 228                                 & 7.61\%                                \\ \hline
\textbf{Health}                         & 190                                 & 6.34\%                                \\ \hline
\textbf{Science}                        & 130                                 & 4.34\%                                \\ \hline
\textbf{Games}                          & 50                                  & 1.67\%                                \\ \hline
\textbf{Business}                       & 35                                  & 1.17\%                                \\ \hline
\textbf{News}                           & 0                                   & 0\%                                   \\ \hline
\textbf{Total}                          & 2996                                & 100\%                                 \\ \hline
\end{tabular}
\end{table}

After finding certain characteristics that help with classifying the URI,
we need to know what percentage of URIs in the Wayback access log have similar characteristics.
First,
we wanted to determine the diversity of the top-level domains (TLDs) in Wayback access log dataset.
Shown in Table \ref{table_dmoz_tld2},
we found that 71.80\% URIs are commercial top-level domain,
.com,
followed by .net,
.org,
and .edu.
This distribution is almost similar to that in DMOZ (Section \ref{tld_dmoz}).

\begin{table}
\caption{Top-level domain analysis for the Wayback Machine server access logs dataset}\label{table_dmoz_tld2}
\begin{tabular}{|l|r|r|}
\hline
\multicolumn{1}{|c|}{\textbf{TLD}}
&\multicolumn{1}{|c|}{\textbf{Num. URIs}}
& \multicolumn{1}{c|}{\textbf{Percent}} \\ \hline
\textbf{com}
&525,651
&71.80\%
\\ \hline

\textbf{net}
&56,589
&7.73\%
 \\ \hline
 
 \textbf{org}
&53,703
&7.34\%
 \\ \hline
 
\textbf{edu}
&8,599
&1.17\%
 \\ \hline
 
\textbf{gov}
&2,343
&0.32\%
 \\ \hline
 
\textbf{us}
&2,071
&0.28\%
 \\ \hline
 
\textbf{others}
&83,174
&11.36\%
\\ \hline

\textbf{Total}
&732,130
& 100\%
  \\ \hline
\end{tabular}
\end{table}

Next,
we want to determine the depth of the requested URIs.
Shown in Table \ref{table_wm_depth} we found that 83.74\% of the URIs in the Wayback access log are depth 0,
essentially top-level web pages.
It means that users often request URIs of depth 0 from the archive.
Since 63.45\% of the correctly classified URIs are of depth 0,
having 83.74\% could enhance the classification results.

\begin{table}
\caption{Depth analysis for Wayback access log dataset}\label{table_wm_depth}
\begin{tabular}{|c|r|r|}
\hline
\textbf{Depth}
& \multicolumn{1}{c|}{\textbf{Count}}
& \multicolumn{1}{c|}{\textbf{Percent}} \\ \hline
\textbf{0} 
&613,121
&83.74\%
\\ \hline
\textbf{1}
&54,008
&7.38\%
\\ \hline
\textbf{2}
&33,644
&4.60\%
 \\ \hline
\textbf{3+}
&31,357
&4.28\%
\\ \hline
\textbf{Total}
&732,130
& 100\%  \\ \hline
\end{tabular}
\end{table}

We saw that having terms found in a dictionary affects classification.
We found that 5.39\% of the Wayback access log URIs contain only words from a dictionary,
and 26.74\% contain at least one word from a dictionary.
These percentages are low and may affect the ability for the requested URI to be correctly classified.

In our DMOZ evaluation, we found that long strings in the domain helped with classification.
When analyzing the Wayback access logs requests, we found that 50.16\% contain long strings in the domain.
We also found that only 3.24\% contain long strings in the path.
In addition,
we found that 12.99\% contain words separated by delimiters in the domain and only 1.54\% in the path.
This also reflects the large percentage of URIs from the access logs with depth 0 (no path).
For classifying most of these requests, we will have to largely rely on domain information.
\section{Conclusions}
In this work we wanted to recommend web pages from a Web archive for a requested ``lost'' URI.
Our work proposes a method to enhance the current response from Web archives when a URI cannot be found (Figure \ref{fig:wm}).
We used both DMOZ and Wikipedia to classify the request and find candidates.
First,
we check if the requested URI is classified in DMOZ or Wikipedia.
If the requested URI is not pre-classified, then  we classify the URI using first-level classification and then deep classification.
This step results in a list of candidates that we filter based on if the web page is archived.
Next we score and rank the candidates based on archival quality,
web page popularity,
temporal similarity,
and URI similarity.

We found that the best method to classify the first-level is using all-grams from the URI  while filtering the TLD and numbers from the URI.
Using a Na{\"i}ve Bayes classifier resulted in a F\textsubscript{1} score of 0.59.
For the second-level classification we measure the accuracy for each classification level.
For second-level classification, the micro-average F\textsubscript{1}=0.30 and for third-level classification, F\textsubscript{1}=0.15.
We also found that 44.89\% of the correctly classified URIs contain a word that exists in a dictionary.
Also,
50.07\% of the correctly classified URIs contain long strings in the domain.
We also analyzed the properties of a sample of URIs requested to the Wayback Machine and found that the large majority were of depth 0,
meaning that our classification will rely largely on domain information.

Future work includes adding other languages,
filtering spam web pages,
and ranking based on how long the web page was not live.
For popularity,
if the access log was saved we can measure how frequently the URI was requested from the archive.
For temporal similarity we can measure the closeness of the creation date of the request and the candidate.

\section{ACKNOWLEDGMENTS}
This work is supported in part by the National Science Foundation,
IIS-1526700.